# Ciphera: A Decentralised Biometric Identity Framework

**Ankit Kanaiyalal Prajapati**
Department of Computer Science and Digital Technologies
University of East London
u2865093@uel.ac.uk

**Shahzad Memon**
Department of Computer Science and Digital Technologies
University of East London
ORCID: 0000-0003-3354-5798

**Mohammed Mahir Rahman**
Department of Computer Science and Digital Technologies
University of East London
ORCID: 0009-0007-4167-1475

**Ameer Al-Nemrat**
Department of Computer Science and Digital Technologies
University of East London
ORCID: 0000-0003-0725-3417

*Abstract*— **Centralised biometric identity systems expose users to single points of failure, opaque verification processes, and irreversible biometric compromise. Decentralised Identifiers (DIDs) and Verifiable Credentials (VCs) offer stronger privacy guarantees, yet their integration with biometric authentication and distributed verification remains insufficiently explored. This paper presents Ciphera, a decentralised biometric identity framework combining privacy-preserving facial recognition, multi-node verification, IPFS-based credential metadata storage, and blockchain-anchored revocation. Evaluated across functional, performance, security, and distributed consistency dimensions, Ciphera achieved an 81% functional success rate, with stable enrolment and authentication but measurable revocation propagation delays and occasional audit-log inconsistencies. Performance testing demonstrated sub-second p95 verification latency of approximately 820ms under concurrent multi-node conditions. Security analysis confirmed strong confidentiality and integrity guarantees, though incomplete liveness detection leaves susceptibility to deepfake and replay attacks. The results demonstrate the feasibility of decentralised biometric identity while identifying key engineering challenges for production-grade deployment.**

***Keywords— Decentralised Identity; Biometric Authentication; Verifiable Credentials; IPFS; Blockchain Revocation; Zero-Knowledge Proof; Multi-Node Verification***

## I. INTRODUCTION

Digital identity systems form the foundation of modern online interactions, enabling authentication, authorisation, and trust establishment across services. Conventional identity infrastructures rely heavily on centralised authorities such as governments, enterprises, and large technology providers that store and manage user data in monolithic repositories. While operationally convenient, such architectures introduce systemic weaknesses including single points of failure, large-scale data breaches, opaque verification processes, and irreversible exposure of biometric information. High-profile compromises of centralised biometric databases demonstrate that once biometric data is leaked, it cannot be revoked or reissued, leaving individuals permanently vulnerable to identity fraud.

Centralised identity systems further lack transparency in how authentication decisions are made and provide limited guarantees regarding retention, propagation, and revocation of identity attributes. Synchronisation delays, administrative bottlenecks, and cross-platform dependencies weaken revocation reliability, limitations that are amplified in multi-service ecosystems where numerous relying parties depend on the same central authority.

Decentralised Identifiers (DIDs) and Verifiable Credentials (VCs), standardised by the W3C, address these issues by enabling cryptographically verifiable, user-controlled identity attributes without reliance on centralised registries [1]. Self-sovereign identity (SSI) literature highlights benefits such as interoperability, reduced institutional dependence, and improved resilience to data breaches [2]. However, practical integration of biometrics into decentralised identity systems remains insufficiently explored. Existing DID deployments predominantly rely on knowledge-based or hardware-token authentication, leaving a clear gap in systems that incorporate biometric identity proofing.

Biometric authentication, particularly facial recognition, offers strong binding between users and their digital identities, supported by advances in deep learning and lightweight models capable of running on edge devices [3]. Yet biometrics introduce unique challenges: they cannot be revoked once compromised, are vulnerable to spoofing and adversarial manipulation, and require privacy-preserving storage and matching mechanisms [4]. When combined with decentralised architectures, additional challenges emerge, including multi-node synchronisation, revocation propagation, audit-log consistency, and distributed verification latency.

To address these gaps, this paper presents Ciphera, a decentralised biometric identity framework integrating privacy-preserving facial recognition, multi-node verification, IPFS-based credential metadata storage, and blockchain-anchored revocation. The system is implemented and evaluated in a controlled distributed environment to provide empirical evidence on the feasibility, performance, and security of decentralised biometric authentication. The primary contribution is the implementation and empirical evaluation of a framework that combines on-device biometric verification, zero-knowledge proof generation, decentralised credential validation, blockchain-anchored revocation, and multi-node distributed verification within a single unified architecture. Unlike existing DID frameworks that focus primarily on decentralised credential management, Ciphera integrates privacy-preserving biometric authentication directly into the decentralised verification lifecycle.

## II. RELATED WORK

Decentralised identity frameworks have been developed to address the limitations of centralised systems that rely on monolithic authorities for identity storage and verification. The W3C DID specification defines identifiers that can be resolved without centralised registries, enabling cryptographically verifiable authentication and user-controlled identity attributes [1]. Verifiable Credentials extend this model by allowing issuers to sign claims that can be independently verified, supporting tamper-evidence and interoperability across distributed environments [2]. SSI literature emphasises reduced institutional dependence, improved privacy guarantees, and enhanced portability of identity attributes [2], [8]. Despite these advantages, challenges persist in revocation, selective disclosure, and metadata leakage.

Biometric authentication has become widely adopted for identity verification due to its strong binding between individuals and digital identities. Advances in deep learning have enabled efficient facial recognition models capable of running on edge devices, improving accuracy and reducing computational overhead [3]. However, biometric systems introduce unique risks, as biometric identifiers cannot be revoked once compromised and templates remain vulnerable to spoofing, adversarial manipulation, and reconstruction attacks [4], [9]. Privacy-preserving mechanisms such as template hashing, cancellable biometrics, and on-device storage have been proposed, though these approaches are rarely evaluated within decentralised identity ecosystems.

Distributed identity verification introduces additional challenges related to synchronisation, revocation propagation, and audit consistency. Revocation in decentralised systems is particularly difficult due to caching delays, asynchronous state updates, and network latency, which can temporarily allow invalid credentials to be accepted [5]. Content-addressed storage systems such as IPFS provide immutability and transparency but do not guarantee real-time consistency across nodes, especially under concurrent access or network congestion [6]. Distributed audit logging typically relies on hash-chain or Merkle-tree structures to ensure tamper-evidence, but inconsistencies in event ordering can compromise verifiability [7]. Foundational work on distributed systems highlights the importance of consistent event ordering and synchronised state replication to maintain system integrity [10]. Blockchain-based consensus mechanisms further support tamper-evident logging and decentralised trust, with Nakamoto consensus providing a widely adopted model for distributed agreement [11].

Although prior research provides strong foundations in decentralised identity, biometrics, and distributed systems, empirical evaluation of systems that combine all three remains limited. Existing studies focus on conceptual models, blockchain identity, or biometric security in isolation, without examining multi-node biometric verification, distributed revocation behaviour, or audit-log consistency in operational settings. Compared with existing frameworks such as Hyperledger Indy and Microsoft ION, Ciphera focuses specifically on the binding between biometric verification and decentralised identity credentials. Hyperledger Indy provides strong support for DIDs, VCs, and ledger-based trust but does not natively define a biometric-to-DID authentication workflow. Microsoft ION supports scalable DID anchoring but similarly focuses on decentralised identifier management rather than biometric identity proofing. Ciphera addresses these gaps by integrating facial embedding generation, distributed biometric verification, IPFS-based metadata storage, and blockchain-anchored revocation into a single operational prototype.

## III. Methodology

This study adopts a design science methodology in which Ciphera is developed as a functional artefact to investigate decentralised biometric identity verification. The methodology encompasses architectural specification, modular implementation, controlled data preparation, and structured testing procedures. The objective was to produce a working prototype capable of supporting biometric-based decentralised authentication under realistic operational conditions, with each architectural decision traceable to a specific security or privacy requirement.

### A. Research Design

The research follows an artefact-centric design process in which Ciphera serves as the primary system under study. The design process involved defining architectural layers, implementing distributed verification components, integrating biometric processing pipelines, and establishing decentralised trust mechanisms. The experimental workflow followed a defined sequence: synthetic biometric samples were first generated and converted into facial embeddings using a TensorFlow Lite pipeline; encrypted embeddings and credential metadata were then submitted through the FastAPI gateway; verifier nodes independently processed biometric similarity matching, IPFS metadata retrieval, and blockchain revocation checks; the gateway aggregated verifier responses and produced the final authentication decision; and audit events were written into a hash-chained log and compared across rapid event sequences to evaluate consistency. This sequence ensured that each result could be traced directly from input data through node processing, gateway aggregation, and final system output.

### B. System Architecture

Ciphera adopts a layered decentralised architecture designed to isolate biometric processing, credential validation, orchestration logic, and trust infrastructure into independently verifiable domains. The architecture minimises centralised trust assumptions by ensuring that raw biometric material, credential secrets, and private cryptographic keys remain under user control throughout the authentication lifecycle. Operational responsibilities are separated across four logical layers: the user-controlled device layer, the orchestration layer, the decentralised verifier layer, and the decentralised trust layer. This separation reduces attack surface concentration while enabling independent verification, modular scaling, and distributed auditability.

The user-controlled device layer is responsible for all sensitive local operations, including biometric capture, facial embedding generation, liveness verification, and zero-knowledge proof construction. No raw biometric material or private key material is transmitted beyond this layer. The orchestration layer is implemented as a FastAPI gateway that handles request validation, verifier coordination, and response aggregation without storing any identity material. The decentralised verifier layer comprises independent nodes that each perform cryptographic proof validation, DID credential verification, and blockchain revocation status checks autonomously. The decentralised trust layer provides the persistent infrastructure, with IPFS serving as content-addressed storage for credential metadata and a blockchain ledger anchoring revocation records immutably.

*Fig. 1* illustrates the complete Ciphera architecture and the interaction between decentralised verification components, orchestration mechanisms, blockchain-backed revocation

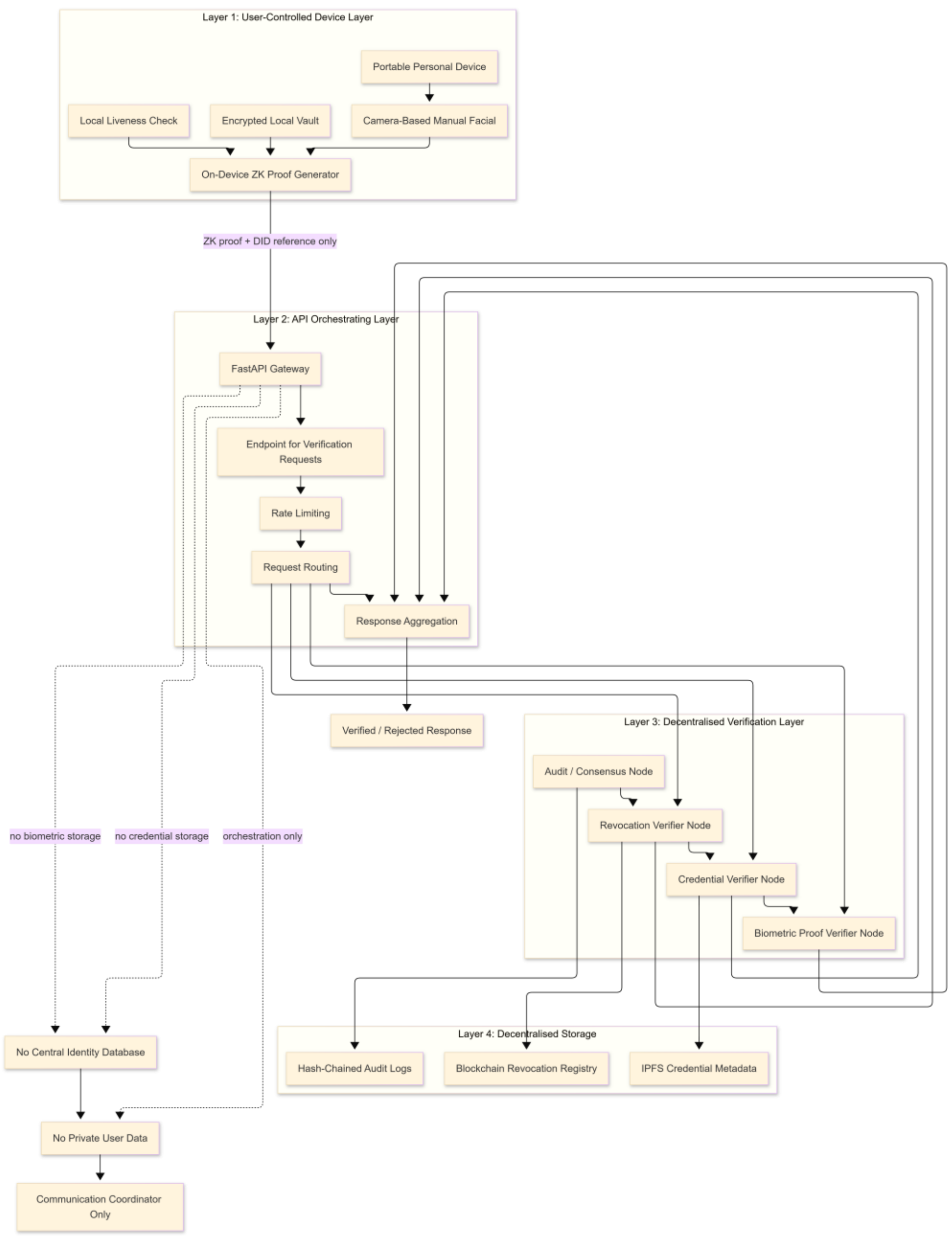

*Figure 1 Complete Layered Ciphera Architecture.*

infrastructure, and privacy-preserving identity verification workflows.

### C. *Authentication Workflow*

Ciphera performs decentralised authentication through a multi-stage verification workflow designed to preserve biometric privacy while maintaining distributed trust validation. Authentication begins on the user-controlled device, where local biometric verification and liveness checks are performed before any external communication occurs. Rather than transmitting raw biometric data, the system generates a cryptographic proof that demonstrates successful local authentication and credential ownership.

Upon submission, the FastAPI gateway validates the request structure and distributes verification tasks to the available verifier nodes. Each node independently validates the zero-knowledge proof, checks the DID credential against the IPFS-stored metadata, and queries the blockchain for the current revocation status of the credential. Node responses are aggregated by the gateway using a consensus threshold, and the final access decision is returned to the client. This multi-node aggregation ensures that no single verifier node can unilaterally grant or deny access, preserving the distributed trust model throughout the authentication lifecycle.

*Fig. 2* illustrates the end-to-end decentralised authentication workflow, from local device verification through gateway orchestration to verifier-node consensus and final access decision.

### D. *Zero-Knowledge Verification Model*

Ciphera incorporates a zero-knowledge verification model to minimise biometric data exposure during authentication. Instead of transmitting raw facial images, biometric templates, or private cryptographic keys, the system generates a cryptographic proof that simultaneously confirms three conditions: the user possesses a valid decentralised credential, local facial verification succeeded, and the credential has not been revoked. This approach ensures that verifier nodes can validate authentication claims without gaining access to the underlying biometric or identity material, significantly reducing the exposure surface compared to conventional biometric verification architectures.

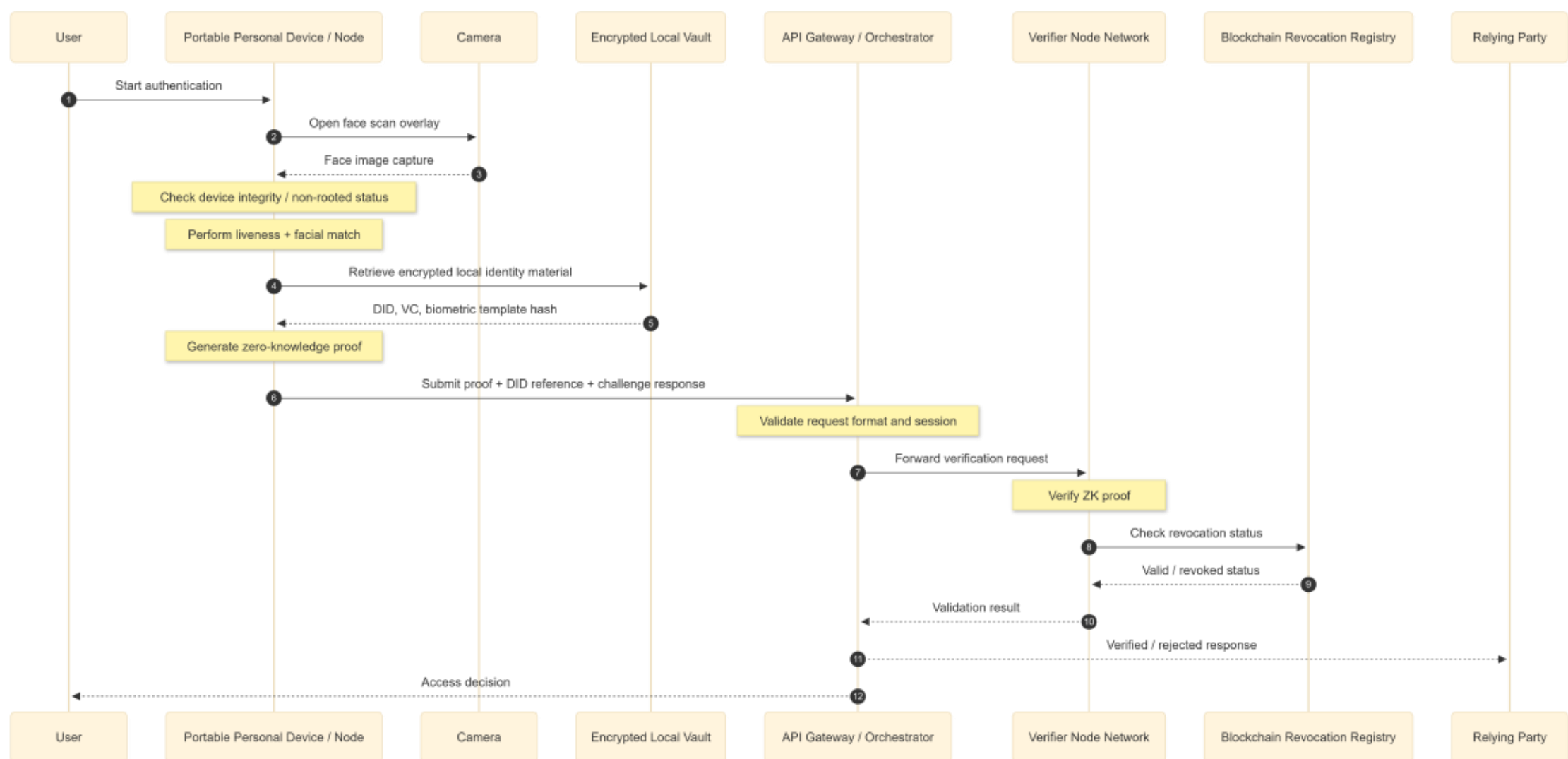

*Figure 2 Decentralised Authentication Workflow.*

*Fig. 3* demonstrates the zero-knowledge verification lifecycle implemented by Ciphera, from local proof generation through distributed proof validation across independent verifier nodes.

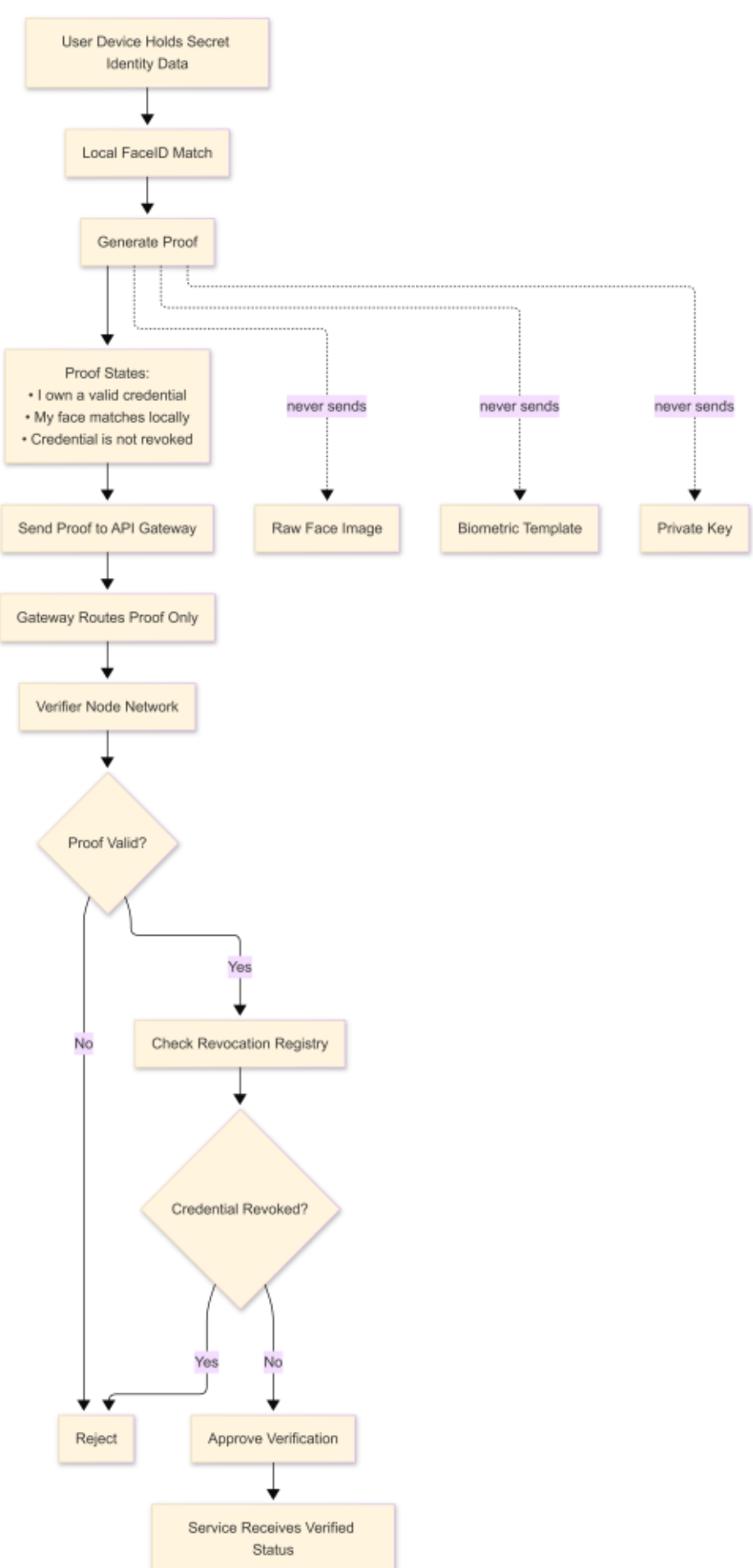

*Figure 3 Zero-Knowledge Verification Lifecycle.*

## *E. Device Integrity and Secure Execution*

Before biometric authentication is permitted, Ciphera validates the integrity and trustworthiness of the executing device. The system performs root and jailbreak detection, secure keystore validation, trusted execution environment checks, and local liveness verification. These checks establish a trusted execution baseline before any proof generation or credential access occurs, ensuring that the security guarantees of the local verification layer cannot be undermined by a compromised device environment. Authentication is blocked if any integrity check fails, preventing the generation of proofs on untrusted hardware.

*Fig. 4* illustrates the device integrity verification process executed prior to biometric authentication, including the conditional flow that blocks proof generation when integrity checks are not satisfied.

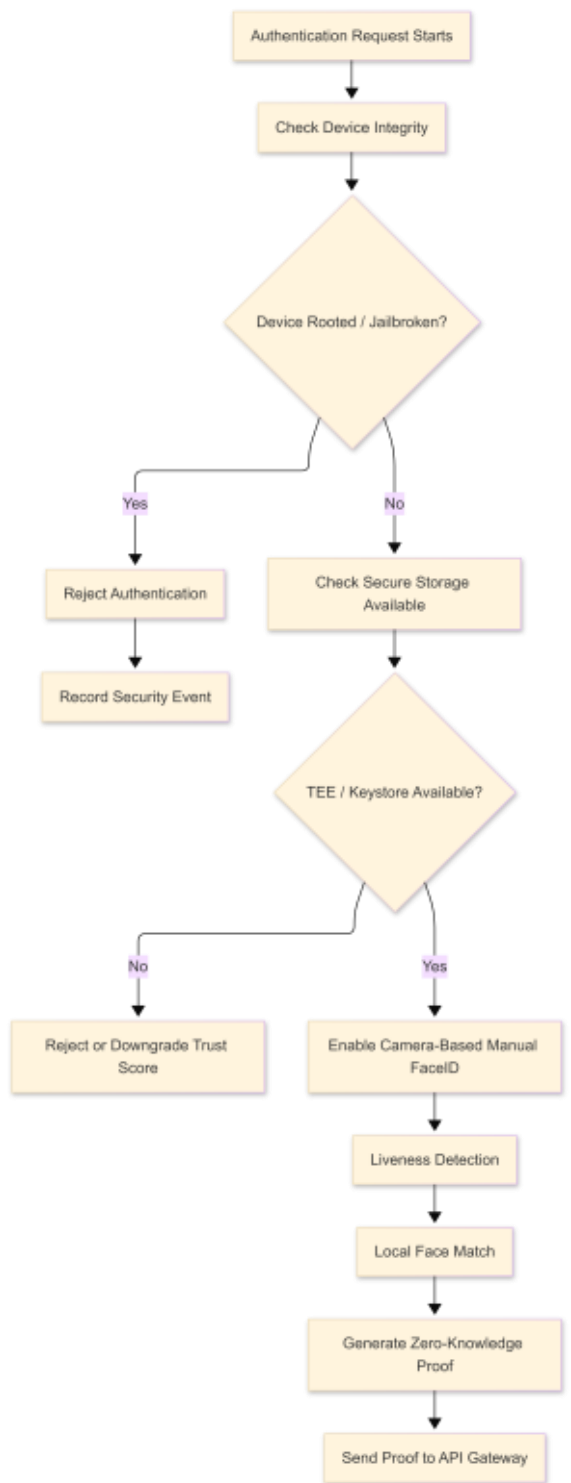

*Figure 4 Device Integrity and Secure Authentication Validation.*

## F. Data Ownership and Privacy Model

Ciphera follows a user-controlled identity ownership model in which sensitive identity assets remain encrypted within the user-controlled environment at all times. Biometric templates, verifiable credentials, and decentralised identity private keys are never stored within centralised infrastructure. The orchestration gateway performs communication coordination only and maintains no persistent identity state. This design ensures that a breach of the gateway or any verifier node does not expose raw biometric material or credential secrets, as these assets never reside outside the user-controlled local vault.

*Fig. 5* demonstrates the privacy-preserving data ownership model, showing the boundary between user-controlled encrypted storage and the orchestration and verification infrastructure.

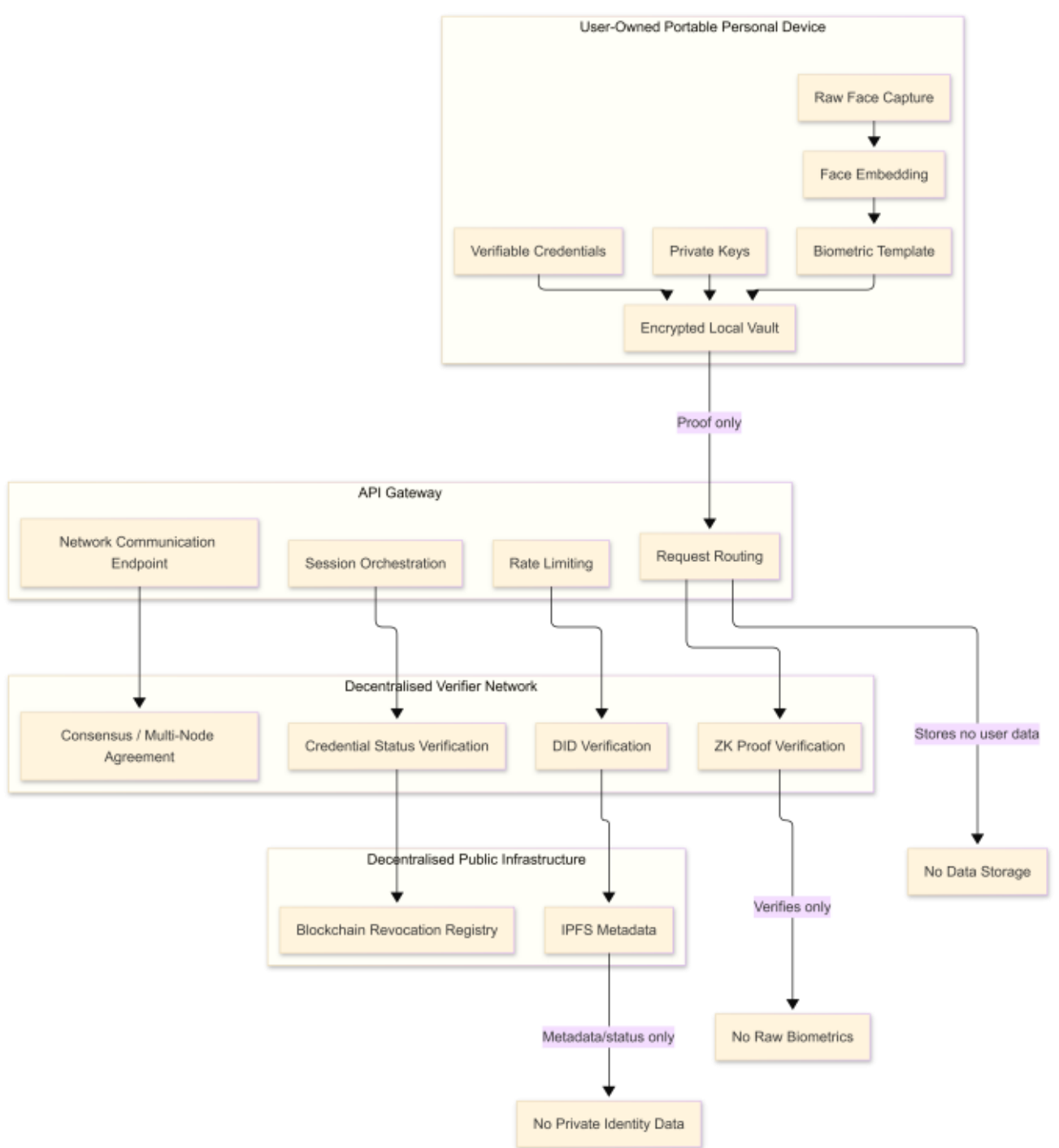

*Figure 5 Privacy-Preserving Data Ownership Model.*

## G. Data Preparation

Data collection involved generating synthetic biometric samples for enrolment and authentication testing. Facial embeddings were produced using the TensorFlow Lite pipeline applied to synthetically generated samples, ensuring no real biometric identities were captured or stored at any stage. Service logs were configured to capture request flow, timestamps, node responses, and audit-log entries to support post-hoc analysis. Load-testing instrumentation was prepared using Locust to measure latency and resource utilisation under controlled concurrency conditions. All biometric samples were anonymised prior to use, and the study was conducted under ethical constraints requiring no collection of real user biometric data.

## H. Testing Procedures

Testing was structured into five categories, each targeting a distinct evaluation dimension. Functional testing verified the correctness of enrolment, authentication, revocation, and audit-log generation workflows across 12 test cases executed over multiple sessions to assess consistency. Performance testing used Locust-driven load tests to measure request latency, throughput, and CPU utilisation under increasing user concurrency, with metrics collected independently at the gateway and each verifier node. Security testing applied the CIA triad, STRIDE threat modelling, and zero-knowledge reasoning to characterise the system's security posture. Test procedures included JWT tampering attempts, unauthorised access trials, and session expiration checks. The threat model is summarised in *Table 1*. Compatibility testing verified consistent behaviour of the web client across Chrome, Firefox, Opera, and Safari, ensuring that the client-side embedding generation pipeline and gateway communication functioned correctly across browser environments. Distributed consistency testing examined synchronisation between verifier nodes by triggering multiple revocation events in rapid succession and observing propagation timing and audit-log ordering behaviour across nodes.

Table 1. STRIDE Threat Model and Ciphera Defence Mechanisms.

| **STRIDE Category** | **Threat Description** | **Ciphera Defence** |
|---|---|---|
| Spoofing | Fake biometric replay | Encrypted embeddings + proposed liveness checks |
| Tampering | Manipulation of templates | SHA-256 hashing + encrypted storage |
| Repudiation | Denial of actions | Hash-chained audit logs |
| Information Disclosure | Data exfiltration | AES-256 encryption + separation of nodes |
| Denial of Service | Flooding Gateway | Rate-limiting + node distribution |
| Elevation of Privilege | Token hijacking | JWT signature validation |

# IV. Results

This section presents the functional, performance, security, compatibility, and distributed consistency results obtained from evaluating the Ciphera prototype in a controlled multi-node environment. All results are derived directly from the implemented system and corresponding test procedures described in Section III.

## A. Functional Evaluation

Functional testing assessed the correctness of enrolment, authentication, revocation, audit-log generation, and session handling across 12 test cases. Ciphera achieved an overall functional success rate of 81%, calculated by averaging pass rates across the five testing categories summarised in *Table 2*. Enrolment achieved a pass rate of 100%, authentication achieved 85%, revocation achieved 80%, audit-log integrity achieved 75%, and session management achieved 70%. Enrolment consistently succeeded and biometric authentication remained stable across repeated trials. Revocation exhibited partial inconsistencies due to propagation delays between verifier nodes, and occasional breaks were observed in audit-log chaining during rapid event sequences. Session management revealed a timeout handling defect that allowed extended access beyond the configured window.

Table 2. Summary of Functional Testing Results.

| Test Category | Pass Rate | Outcome Summary |
|---|---|---|
| Enrolment | 100% | Enrolment process completed successfully across all cases; biometric templates stored correctly. |
| Authentication | 83% | Authentication generally successful; minor delays observed during face-match under load. |
| Revocation | 80% | Revocation recorded on blockchain and IPFS; propagation delay between Node 1 and Node 2 detected. |
| Audit Log Integrity | 75% | Hash-chain mostly consistent; occasional breaks observed during rapid revocation events. |
| Session Management | 70% | JWT validation correct; session timeout bug allowed extended access beyond configured window. |

## B. Performance Evaluation

Performance testing measured latency and resource utilisation under increasing user concurrency. The system achieved a p95 latency of approximately 820ms, indicating that multi-node verification and gateway orchestration remained within acceptable bounds for interactive authentication workflows. CPU utilisation remained below 50% across all nodes, demonstrating computational efficiency and headroom for scaling. Latency distribution patterns confirmed that biometric embedding generation and IPFS retrieval were the dominant contributors to overall response time.

## C. Security Evaluation

Security analysis was conducted using the CIA triad, STRIDE threat modelling, and zero-knowledge reasoning. Ciphera maintained confidentiality through AES-256 encryption of biometric embeddings and strict JWT-based session validation. Integrity was supported by SHA-256 hashing of templates, blockchain-anchored revocation events, and hash-chained audit logs. Availability was reinforced by distributing verification responsibilities across independent nodes. The STRIDE assessment identified spoofing and denial-of-service as the highest-risk categories. While the system incorporates encrypted embeddings and proposes liveness checks, the absence of a fully implemented liveness module leaves susceptibility to deepfake and replay attacks. Rate limiting and node distribution mitigated basic flooding attempts, though automated failover and inter-node TLS were not implemented, limiting resilience under adversarial load.

## D. Compatibility Evaluation

Compatibility testing confirmed consistent behaviour across Chrome, Firefox, Opera, and Safari. All browsers successfully executed the client-side embedding generation pipeline and maintained stable communication with the gateway. Minor UI rendering differences were observed but did not affect authentication flow or data integrity.

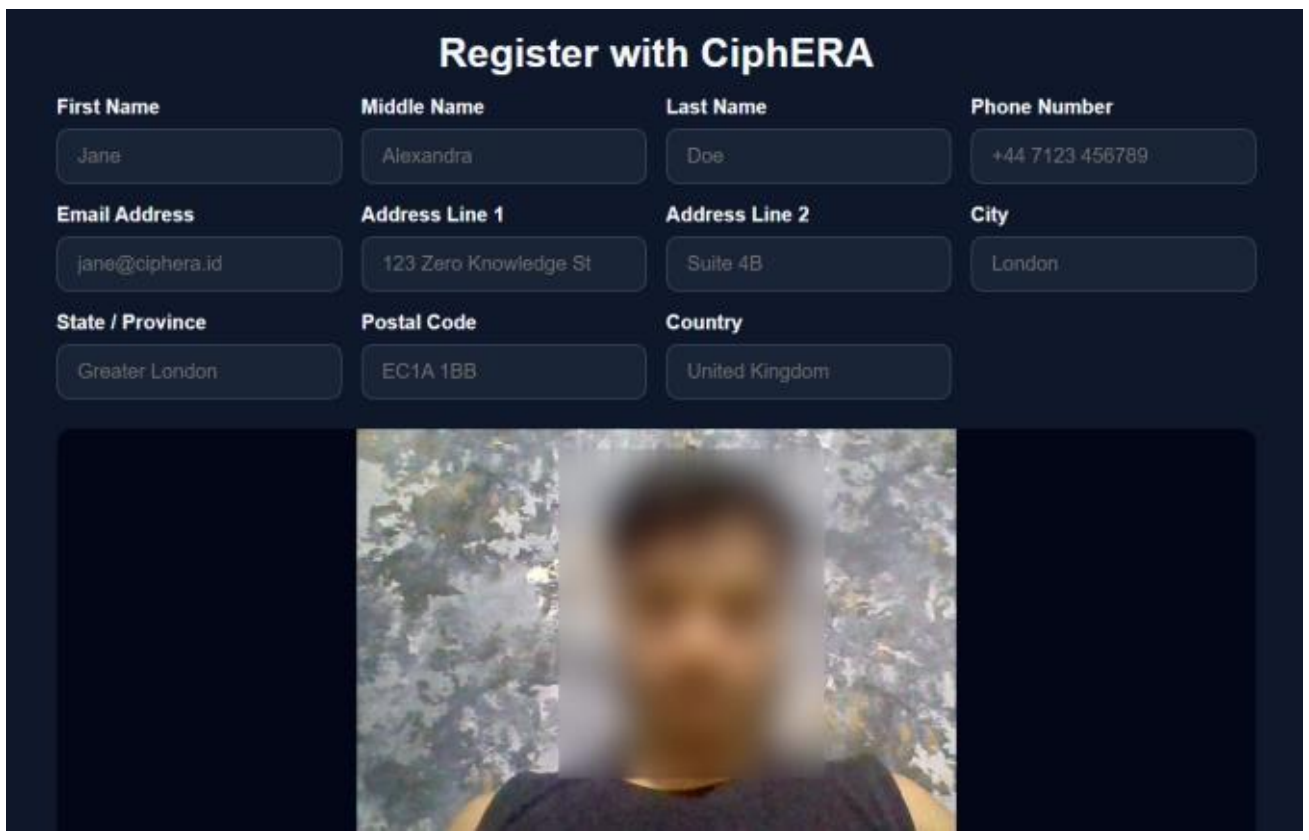

*Figure 6 User Biometric Registration Interface.*

*Fig. 6* represents the initial enrolment stage of the Ciphera system, where users register their identity and facial biometrics by providing basic profile information alongside multiple facial samples captured through the camera interface.

## E. Distributed Consistency Evaluation

Distributed consistency testing focused on revocation propagation and audit-log ordering. Revocation events were correctly recorded on the blockchain and reflected in IPFS metadata; however, verifier nodes exhibited temporary inconsistencies due to caching delays. During rapid revocation sequences, hash-chain audit logs occasionally diverged, indicating sensitivity to event ordering and concurrency. These inconsistencies occurred primarily when multiple distributed events were generated before verifier nodes reached synchronised state agreement, a behaviour attributable to asynchronous event propagation, cache invalidation timing, race conditions, and non-deterministic log ordering across nodes. These findings highlight the need for stronger synchronisation mechanisms and deterministic log-writing policies.

## F. Comparative Analysis

*Table 3* positions Ciphera against four existing identity frameworks across six evaluation dimensions. Ciphera is the only system in the comparison set that combines facial-embedding-based biometric verification, full multi-node decentralisation, DID and VC-based public key infrastructure, blockchain revocation, and zero-knowledge proof support, while maintaining low centralised storage risk. Hyperledger Indy and Sovrin offer partial zero-knowledge support within distributed trust architectures but do not define native biometric verification workflows. Microsoft ION provides scalable DID anchoring without biometric verification or zero-knowledge support. MOSIP supports biometric identity but operates within a semi-centralised architecture with a central revocation authority and high centralised storage risk.

Table 3. Comparison of Decentralised Identity Frameworks

| System | Biometric Verification | Decentralization | Public Key Infrastructure | Revocation | Zero-Knowledge Support | Centralized Storage Risk |
|---|---|---|---|---|---|---|
| Ciphera | Facial embeddings + local matching | Multi-node distributed | DID + VC | Blockchain revocation | Yes | Low |
| Hyperledger Indy | Credential-based | Distributed ledger | DID framework | Ledger revocation | Partial | Medium |
| Microsoft ION | DID anchoring | Sidetree-based | DID framework | External handling | No | Medium |
| Sovrin | SSI ecosystem | Distributed trust network | DID infrastructure | Ledger-based | Partial | Medium |
| MOSIP | Biometric identity | Semi-centralized | PKI-based | Central authority | No | High |

## V. Discussion

The findings demonstrate that decentralised biometric identity verification is technically feasible when biometric processing, credential validation, and revocation verification are separated into independent verification domains. The layered architecture reduced centralised trust dependency and improved privacy preservation by ensuring that raw biometric data and private identity material never leave the user-controlled environment. The 81% functional success rate, combined with sub-second p95 latency under concurrent multi-node conditions, confirms that the architectural separation of gateway and verifier nodes is both operationally viable and computationally efficient.

However, the evaluation also revealed several limitations associated with distributed synchronisation and decentralised revocation consistency. Revocation latency remained dependent on verifier-node synchronisation intervals, cache invalidation timing, and blockchain propagation delays. This introduces a temporary trust window during which outdated credential states may persist across nodes, a known challenge in distributed revocation systems where asynchronous state updates can temporarily allow invalid credentials to be accepted [5]. Addressing this will require deterministic cache invalidation strategies and event-driven synchronisation mechanisms that reduce the window between a revocation event and its reflection across all verifier nodes.

Audit-log inconsistencies observed during rapid revocation sequences similarly point to a broader challenge in maintaining deterministic event ordering across distributed nodes. Hash-chained logging provides strong tamper-evidence when event ordering is consistent, but the sensitivity of the current implementation to concurrency and race conditions limits its reliability under high-frequency operational scenarios. Stricter log-writing policies, combined with distributed ordering protocols drawing on established work in consistent event sequencing [10], represent a viable path toward resolving this limitation.

The security analysis further demonstrated that decentralised verification alone does not eliminate biometric spoofing risk. Although Ciphera reduces centralised biometric exposure through local embedding generation and zero-knowledge verification, the absence of production-grade liveness detection leaves the system vulnerable to replay and deepfake attacks, currently the most significant unresolved security gap in the prototype. Integrating a robust liveness detection pipeline before biometric embeddings are accepted for verification is therefore an essential requirement for any production deployment. This is particularly critical given the growing accessibility of deepfake generation tools, which lower the barrier for presentation attacks against facial recognition systems.

The use of synthetic biometric samples and controlled testing conditions represents a further limitation on the generalisability of these findings. Real user populations introduce greater variation in lighting, facial orientation, camera quality, network conditions, and adversarial behaviour than was modelled in the controlled evaluation environment. Large-scale deployment would also surface additional challenges in verifier-node scaling, inter-node TLS, and automated failover that the current prototype does not address. These constraints do not undermine the core feasibility findings but indicate that additional engineering work is required before Ciphera can be considered production-ready.

Taken together, the results position Ciphera as a meaningful practical step toward decentralised biometric identity, demonstrating that the integration of privacy-preserving facial recognition, IPFS-based credential metadata, blockchain-anchored revocation, and multi-node verification is achievable within a unified operational architecture. The identified limitations provide a concrete engineering agenda for future work rather than fundamental objections to the decentralised biometric identity model itself.

## VI. Conclusion

This paper addressed the limitations of centralised biometric identity systems, which suffer from single points of failure, opaque verification processes, and irreversible exposure of biometric data. Existing decentralised identity frameworks provide strong theoretical foundations through DIDs and Verifiable Credentials, yet prior work has not demonstrated how biometric authentication can be practically integrated into a distributed, multi-node verification environment. Ciphera contributes a working implementation that combines privacy-preserving facial recognition, IPFS-based credential metadata storage, blockchain-anchored revocation, and independent verifier nodes within a unified operational architecture.

The evaluation confirms that decentralised biometric authentication is feasible and that the architectural separation of gateway, biometric, and credential nodes supports secure and efficient operation. Functional testing achieved an 81% success rate, with stable enrolment and authentication pipelines, while performance results demonstrated sub-second p95 latency under concurrent multi-node conditions. These results provide empirical evidence that the design principles underlying Ciphera are sound and operationally viable.

At the same time, the evaluation exposes several engineering challenges that remain unresolved. Revocation propagation delays highlight the difficulty of maintaining consistent credential states across distributed nodes. Audit-log inconsistencies during rapid event sequences demonstrate the sensitivity of hash-chained logging to concurrency and non-deterministic event ordering. The absence of a complete liveness detection module represents the most significant security gap, leaving the system vulnerable to deepfake and replay attacks.

Future work will focus on strengthening revocation synchronisation through deterministic cache invalidation and event-driven update mechanisms. Enhancing audit-log consistency will require stricter ordering guarantees and improved inter-node coordination. Implementing a robust liveness detection pipeline is essential to mitigate spoofing and deepfake threats. Additional work will explore automated failover, inter-node TLS, and scaling the architecture to larger verifier networks. These improvements will support the development of more resilient, secure, and scalable decentralised biometric identity systems.

## Acknowledgment

The author would like to thank the academic supervisors and faculty members whose guidance and feedback supported the development of this work.